\documentclass[twocolumn,prl,showpacs,preprintnumbers,amsmath,amssymb]{revtex4}

\usepackage{graphicx}
\usepackage{dcolumn}
\usepackage{bm}

\begin{document}
\preprint{APS/123-QED}
\author{T.~Cichorek,$^{1,2}$ A.~Sanchez,$^{1}$ P.~Gegenwart,$^{1}$ F.~Weickert,$^{1}$ A.~Wojakowski,$^{2}$ Z.~Henkie,$^{2}$ G.~Auffermann,$^{1}$
S.~Paschen,$^{1}$ R.~Kniep,$^{1}$ and F.~Steglich$^{1}$}

\address{$^{1}$ Max Planck Institute for Chemical Physics
of Solids, 01187 Dresden, Germany
\\ $^{2}$Institute of Low Temperature and Structure Research, Polish Academy of Sciences, 50-950 Wroclaw, Poland}
\title{Two-Channel Kondo Effect in Glass-Like ThAsSe}

\date{\today}

\begin{abstract}
We present low-temperature heat and charge transport as well as
caloric properties of a ThAsSe single crystal. An extra
--\textit{A}\textit{T}$^{1/2}$ term in the electrical resistivity,
independent of magnetic fields as high as 14~T, provides evidence
for an unusual scattering of conduction electrons. Additionally,
both the thermal conductivity and specific heat show a glass-type
temperature dependence which signifies the presence of tunneling
states. These observations apparently point to an experimental
realization of a two-channel Kondo effect derived from structural
two-level systems.
\end{abstract}

\maketitle

Since the pioneering work by Nozi\'{e}res and Blandin \cite{Noz},
the search for a two-channel Kondo (2CK) effect has been the
subject of considerable scientific interest. However, its
experimental realization requires a strict channel symmetry that
is hard to achieve in the spin Kondo problem. As a viable
alternative, one considers an equivalent Kondo effect that
originates from scattering centers with orbital instead of spin
degrees of freedom like, e.g., the quadrupole momenta in certain
heavy-fermion systems \cite{Cox}. In this so-called orbital Kondo
effect, the spin of the conduction electrons plays the role of a
silent channel index. The influence of a magnetic field on the
orbital Kondo resonance must be very weak and, therefore,
distinctly different from the destruction of the spin Kondo
resonance by a comparatively low magnetic field \cite{Zhu}.

As originally suggested by Zawadowski\textit{ et al.}
\cite{Zaw,Vladar}, the interaction between structural two-level
systems (TLS) and the conduction electrons may also lead to the
orbital Kondo effect and hence to the 2CK problem. In the TLS
Kondo model, a single tunneling center (e.g., an atom that
quantum-mechanically tunnels between two minima of a double-well
potential) immersed in the Fermi sea is expected to behave like a
spin$-$$\frac{1}{2}$ impurity coupled to the conduction electrons
\cite{Coch,Kondo}. Since this is essentially a one-impurity Kondo
problem, a considerably stronger effect on the transport
properties rather than on the thermodynamic ones is anticipated
\cite{Cox}. On the other hand, it is still debatable whether the
strong-coupling 2CK fixed point due to the TLS can be
experimentally achieved or not. This is mainly because a Kondo
temperature \textit{T}$_{K}$ as small as 10$^{-3}$--10$^{-2}$~K is
expected \cite{Cox}. In addition, the contribution of higher
excited states may reduce \textit{T}$_{K}$ by up to three orders
of magnitude to negligibly small values \cite{Aleiner1, Aleiner2}.
Furthermore, even if the second vibrational level of the tunneling
atom is just above the potential barrier, for an intermediate
heavy atom a \textit{T}$_{K}$ value below 0.5~K is predicted
\cite{Borda}. Nevertheless, various experimental observations
obtained at liquid-helium temperatures are thought to be caused by
an interaction between the conduction electrons and TLS
\cite{Aliev, Halb, Ralph, Csonka}. For example, an orbital Kondo
problem with large \textit{T}$_{K}$ value is inferred from the
width of the Kondo resonance on the Cr(001) surface investigated
by scanning tunneling microscopy \cite{Kole}. Moreover, recent
theoretical studies have shown that not all of the possible
internal structures of the defect have been explored
\cite{Miyake}. Finally, a significant enhancement of the Kondo
temperature is suggested if the conduction electrons are coupled
to the tunneling impurity via resonant scattering \cite{Zarand}.

Recent work on the diamagnetic compound ThAsSe has demonstrated
the importance of tunneling states to the charge transport on a
macroscopic scale \cite{cich1}: At temperatures lower than about
20~K, an additional term in the resistivity $\rho$(\textit{T}) was
detected that frequently exhibited a complex temperature
dependence. For other single crystals of ThAsSe, however, a
simpler, i.e., a logarithmic increase of $\rho$(\textit{T}) upon
cooling, followed by a saturation below
\textit{T}$_{S}$$\simeq$0.2--2~K, was found \cite{cich1}. These
peculiarities, together with their independence on both strong
magnetic fields and high hydrostatic pressures, point to a Kondo
effect derived from structural TLS. Furthermore, some tendency of
$\rho$(\textit{T}) to pass from a logarithmic to a
--\textit{A}\textit{T}$^{1/2}$ behavior was observed in a ThAsSe
sample with \textit{T}$_{S}$$\simeq$0.2~K, hinting at the
development of a possible 2CK state \cite{cich2}. A crude
comparison of the experimental results with the theoretical ones
yielded a Kondo temperature of 4--5 K \cite{cich2}. The presence
of tunneling centers in single-crystalline ThAsSe was directly
reflected by a quasilinear-in-\textit{T} term of
\textit{nonelectronic} origin in the low-\textit{T} specific heat.
TLS centers are apparently located in the As-Se substructure, as
suggested by x-ray and transmission-electron-microscopy studies
\cite{Henkie} as well as $^{77}$Se NMR measurements \cite{Mich}.

This Letter reports the observation of a 2CK state originating
from a scattering of the conduction electrons off structural TLS
in ThAsSe. All experiments have been performed on the same single
crystal. The synthesis procedure is described in Ref.
\cite{Henkie}. Due to the platelike shape of the crystal, its
transport properties have been investigated in the \textit{ab}
plane only. The thermal conductivity was measured in a
$^{3}$He--$^{4}$He dilution refrigerator utilizing a steady-state
method. The electrical resistivity was studied in zero and applied
magnetic fields up to 14~T using different equipment below and
above 4~K. The specific heat was measured in a $^{3}$He cryostat
with the aid of a thermal-relaxation method.

It is well established that TLS determine the low-temperature
thermal properties of matter with some kind of disorder, leading
to anomalous terms in the temperature dependences of both the
thermal conductivity and the specific heat of comparable
magnitude. The low-\textit{T} thermal properties of ThAsSe are
remarkably similar to those of amorphous solids: Fig. 1(a) shows
the temperature dependence of the thermal conductivity
$\kappa$(\textit{T}) of ThAsSe. The electronic contribution
$\kappa_{el}^{WF}$ was estimated from the electrical resistivity
using the Wiedemann-Franz law. Since for \textit{T}$\lesssim$1~K,
$\kappa_{el}^{WF}$ (dotted line) is distinctly smaller than the
measured total $\kappa$(\textit{T}), the heat transport is
dominantly carried by phonons. This becomes also evident by
comparing the $\kappa$(\textit{T}) data of ThAsSe with the
published ones for the dielectric glass As$_{2}$S$_{3}$
\cite{Zeller, Stephens} $-$ a prototype of
pnictogenchalcogenide-based TLS materials. For \textit{T}$<$0.7~K,
$\kappa$(\textit{T}) of As$_{2}$S$_{3}$ follows the relation
$\kappa$(\textit{T})=\textit{C}(\textit{T}/$\alpha$)$^{\delta}$
with $\alpha$=1~K, $\delta$=1.92 and
\textit{C}=17$\times$10$^{-4}$~W/Kcm \cite{Stephens}. A universal
power-law dependence of $\kappa$(\textit{T}) with
$\delta$=1.9$\pm$0.1 is characteristic for systems whose
dominating phonon thermal conductivity is limited by scattering
from TLS. For ThAsSe, $\kappa$(\textit{T}) can be well described,
below 0.8~K, by the same relation with $\alpha$=1~K, $\delta$=1.97
and \textit{C}=23$\times$10$^{-4}$~W/Kcm (cf. solid line). While
the exponent $\delta$ fits perfectly into the narrow range
1.9$\pm$0.1, the value of \textit{C} is only slightly larger than
that for the vitreous As$_{2}$S$_{3}$ (part of this difference may
be well attributed to errors in the estimation of the geometric
form factor of the ThAsSe crystal). Furthermore, phonon scattering
off conduction electrons, leading also to a
\textit{T}$^{2}$-contribution to $\kappa$(\textit{T}), seems to be
negligible in ThAsSe: the product of the mean free path of the
charge carriers, \textit{l}$_{c}$, and the wave number of the
dominating phonon, \textit{q}$_{ph}$, is estimated to be 0.5,
i.e., below the so-called Pippard ineffectiveness condition
\textit{l}$_{c}$\textit{q}$_{ph}$=1 \cite{Pippard}. Thus, the
low-\textit{T} thermal conductivity provides striking evidence for
the TLS being the dominating scattering centers for the
propagating phonons in ThAsSe.

Additional evidence for tunneling centers in ThAsSe comes from the
low-\textit{T} specific heat results in Fig. 1(b). Here, we
compare \textit{C}(\textit{T}) of our ThAsSe crystal with
\textit{C}(\textit{T}) previously obtained for an ensemble of
small single-crystalline pieces of this material \cite{cich1}.
Though quantitatively slightly different, the two sets of data
confirm the existence of an anomalous term, $\sim$\textit{T}, to
the specific heat. At low temperatures, this term adds to the
specific heat due to charge carriers and phonons,
$\gamma$\textit{T}+$\beta$\textit{T}$^{3}$, determined at
1.7~K$<$\textit{T}$<$5 K. The additional low-\textit{T}
contributions is ascribed to the presence of TLS.

\begin{figure}
\includegraphics[width=0.45\textwidth]{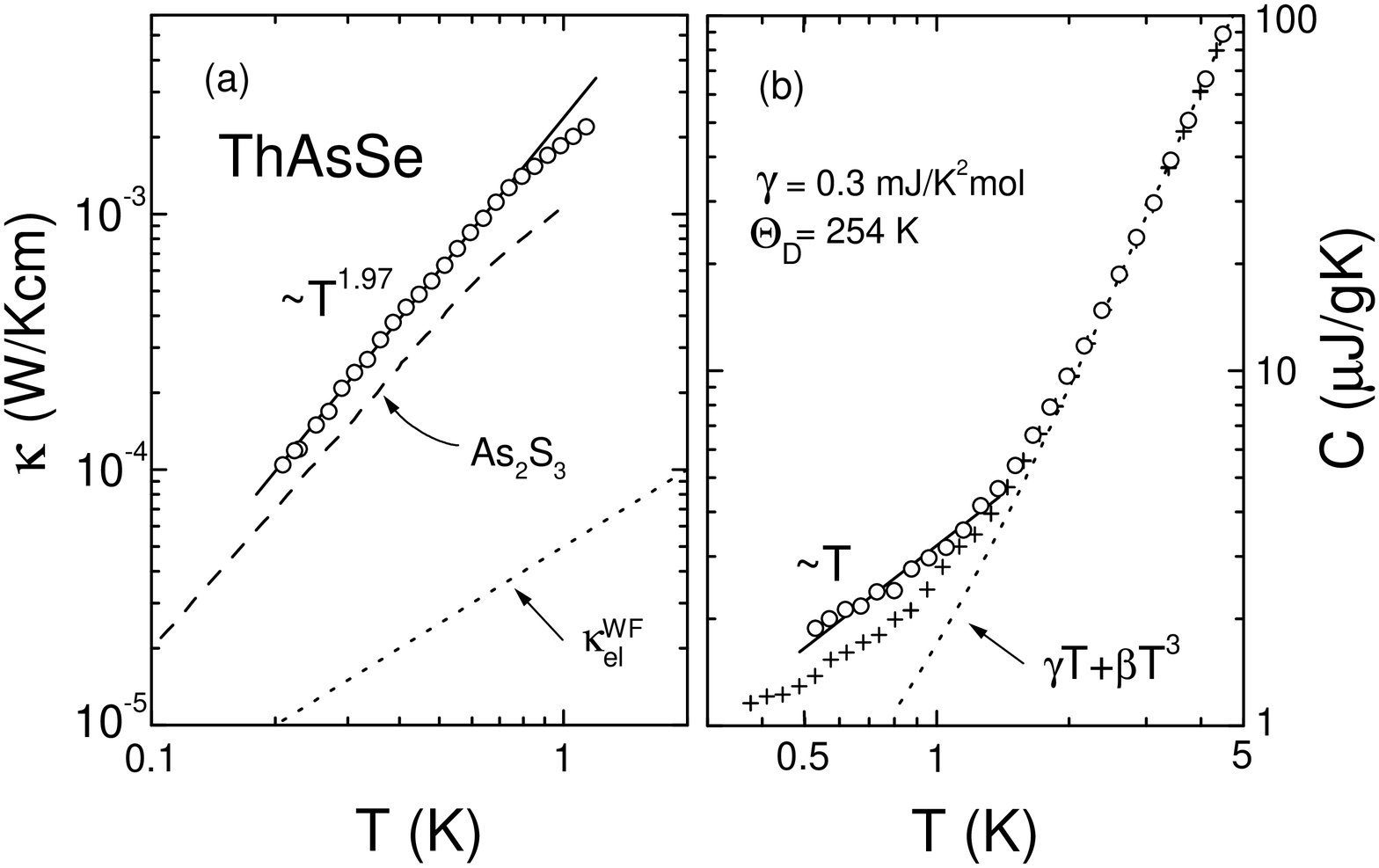}
\caption{\label{fig:epsart} Low-temperature thermal properties of
the ThAsSe single crystal. (a): Total thermal conductivity
(circles) and estimated electronic contribution $\kappa_{el}^{WF}$
(dotted line) as a function of temperature. The solid line is a
fit to the data as described in the text. For comparison,
$\kappa$(\textit{T}) of the noncrystalline insulator
As$_{2}$S$_{3}$ is also shown \cite{Stephens}. (b): Specific heat
in a double-logarithmic plot (circles). Crosses are the previous
\textit{C}(\textit{T}) data for a sample composed of several
pieces \cite{cich1}. The dotted line shows the
$\gamma$\textit{T}+$\beta$\textit{T}$^{3}$ dependence with
$\gamma$=0.3~mJ/K$^{2}$mol, the Sommerfeld coefficient of the
electronic specific heat, and
$\beta$=3$\times$1944/$\Theta_{D}^{3}$ in units of J/K$^{4}$mol.
The parameters $\gamma$ and $\beta$ were determined from the fit
in the temperature range 1.7-5~K. The solid line indicates the
presence of an additional linear-in-\textit{T} term due to TLS.}
\end{figure}

Having established the glassy character of the ThAsSe single
crystal studied, we now turn to the discussion of its electronic
transport properties at low temperatures. As shown in Fig. 2, an
additional contribution to $\rho$(\textit{T}) emerges at
temperatures below 16 K. Here we plotted the relative change of
the resistivity normalized to the corresponding value at 1~K,
$\Delta\rho/\rho_{1 \textrm{K}}$. For \textit{T}$\lesssim$0.9~K
and \textit{B}=0, the resistivity is seen to level off. By
applying a magnetic field larger than \textit{B}=1~T,
$\rho$(\textit{T}) depends strictly linearly on \textit{T}$^{1/2}$
in a wide temperature window, i.e., from around 0.16~K to above
12~K at \textit{B}=14~T. The coefficient of the
--\textit{A}\textit{T}$^{1/2}$ term amounts to
\textit{A}=~0.38~$\mu\Omega$cm/K$^{1/2}$ for all fields
\textit{B}$\geq$1~T. [Here, we assumed
$\rho_{ab}$(300~K)=220~$\mu\Omega$cm \cite{Schoe}, see also Fig.
3(c).] Finally, we note that, while this anomalous contribution to
the resistivity sets in at slightly lower temperatures with
increasing \textit{B} (cf. arrows in Fig.~2), $\rho$(\textit{T})
begins to saturate -- independently of \textit{B} -- at the same
temperature below 0.1~K \cite{heating}.

The field-independent \textit{T}$^{1/2}$ increase  of
$\rho$(\textit{T}) observed upon lowering the temperature can
neither be attributed to weak localization \cite{Lee} nor to
electron-electron interactions in a three-dimensional disordered
system \cite{Alts}. In fact, both type of quantum corrections to
the resistivity are highly sensitive to magnetic fields, which
holds true even in the presence of strong spin-orbit coupling
\cite{Lee, Alts}. For example \cite{Bergmann}, the interference of
the wavefunctions of the electrons moving along a closed loop is
weakened or even destroyed by a magnetic field of the order of
tens of an Oe only. We, therefore, attribute the field-independent
--\textit{AT}$^{1/2}$ term in the electrical resistivity to
electron scattering off the TLS, whose existence had been
estimated by the thermal measurements described above. This is, to
our knowledge, the first-ever observation of the 2CK state
originating from TLS in a macroscopic system.

Within the scope of electron-TLS interaction, an increase of the
zero-field resistivity upon cooling, being weaker than
$-$\textit{A}\textit{T}$^{1/2}$ as measured for \textit{B}$\neq$0
(Fig. 2), is somewhat surprising. This holds true also for the
strong field dependence of $\rho$(\textit{T}) at the lowest
temperatures for \textit{B}$<$1~T (Fig.~2): while some deviations
from \textit{A}=0.38~$\mu\Omega$cm/K$^{1/2}$ are still visible in
\textit{B}=0.2~T below around 1.3~K, already \textit{B}=1~T acts
in the same way as 7 and 14~T do. Therefore, we suppose that the
2CK effect in the weak field limit is hidden by another phenomenon
being suppressed already by \textit{B}$\simeq$1~T. To explore this
possibility further, the isothermal response of the resistivity to
a magnetic field was studied.

\begin{figure}
\includegraphics[width=0.35\textwidth]{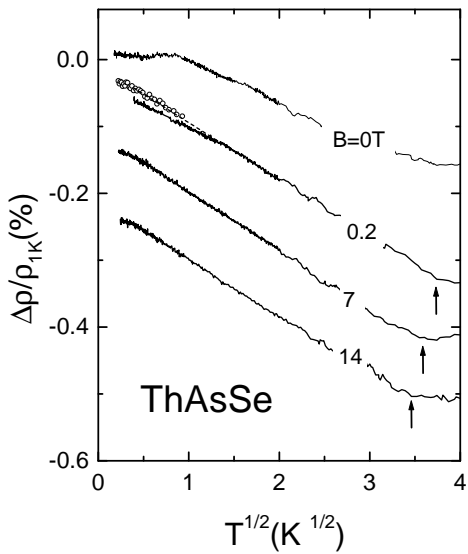}
\caption{\label{fig:epsart} Low-temperature electrical resistivity
of ThAsSe as $\Delta$$\rho$/$\rho$$_{1 \textrm{K}}$ vs
\textit{T}$^{1/2}$ in varying magnetic fields applied along the
\textit{c} axis. For clarity, the different curves in
\textit{B}$>$0 were shifted subsequently by 0.1\%. Circles display
the \textit{B}=1 T data, and dashed line represents a
--\textit{AT}$^{1/2}$ behavior with
\textit{A}=0.38~$\mu\Omega$cm/K$^{1/2}$, independent of field at
\textit{B}$\geq$1 T. Arrows indicate the upper limit of the
square-root temperature dependence of the resistivity.}
\end{figure}

The magnetoresistivity (MR) data for ThAsSe are shown in Fig.~3(a)
as ($\rho_{B}$-$\rho_{0}$)/$\rho_{0}$ vs \textit{B}$^{1/2}$. The
measurements were performed in the temperature window
0.1~K$\leq$\textit{T}$\leq$10~K where the --\textit{AT}$^{1/2}$
dependence was observed in $\rho$(\textit{T}). A positive MR whose
magnitude gradually decreases with increasing temperature is found
at \textit{B}$<$1~T only. At \textit{T}=10~K,
($\rho_{B}$-$\rho_{0}$)/$\rho_{0}$ is practically zero in this
field range. This low-field effect hints at some quantum
corrections. Most probably, the positive magnetoresistivity
reflects spin-orbit scattering that rotates the spin of the
conduction electrons and yields a destructive interference of the
electron wavefunctions \cite{Bergmann}. Consequently, the
differences between the results obtained at \textit{B}$\leq$0.2~T
and those obtained at \textit{B}$\geq$1~T, as depicted in Fig. 2,
are tentatively ascribed to quantum corrections.

Further arguments against the interference effects in higher
fields derive from the data at \textit{B}$\geq$1~T, where the MR
is negative. Because of the \textit{T}-independent slope of
($\rho_{B}$-$\rho_{0}$)/$\rho_{0}$ vs \textit{B} curves displayed
in Fig. 3(a) for \textit{B}$\geq$1~T, this negative MR cannot be
ascribed to weak localization \cite{Bergmann}. Furthermore, at
fields of the order of
(2$\mu_{B}$/\textit{k}$_{B}$\textit{T})$^{-1}$, one should expect
a deviation from the observed \textit{B}$^{1/2}$ behavior due to
the electron-electron interaction \cite{Lee}, which is not
resolved in the data of Fig. 3(a).

\begin{figure}
\includegraphics[width=0.48\textwidth]{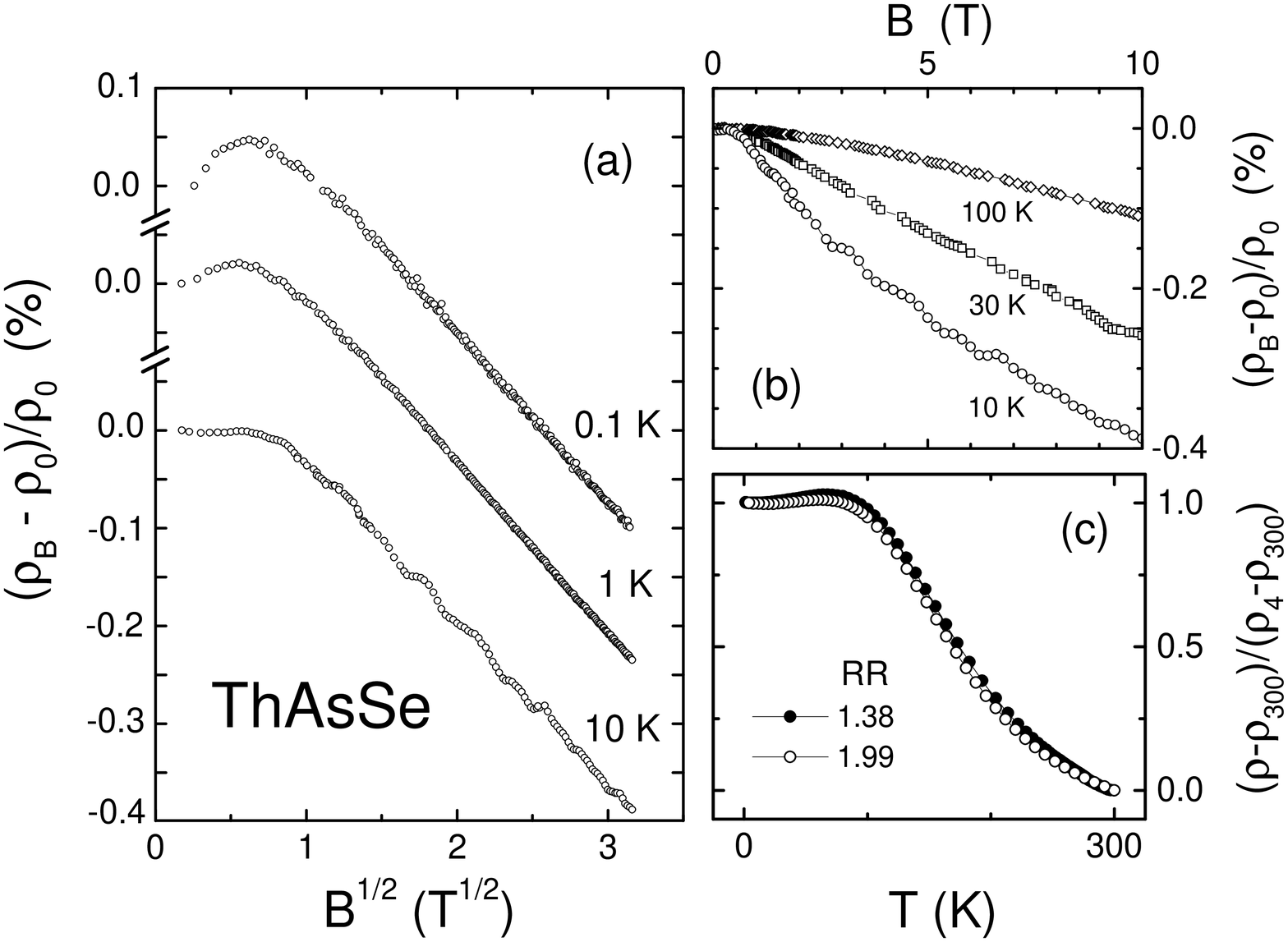}
\caption{\label{fig:epsart} (a) Magnetic field dependence of the
electrical resistivity for ThAsSe at temperatures where a
-\textit{AT}$^{1/2}$ term is observed (see Fig. 2). (b)
Magnetoresistivity at higher temperatures. (c) The normalized
resistivity for two single crystals of ThAsSe with very different
resistivity ratios RR=$\rho_{4 \textrm{K}}/\rho_{300
\textrm{K}}$.}
\end{figure}

A small and negative MR appears to be characteristic for ThAsSe
[Fig. 3(b)]. However, the MR decreasing as \textit{B}$^{1/2}$ has
been observed only at \textit{T}$\leq$10~K. Additionally, the
slope of this isothermal \textit{B}$^{1/2}$ dependence accounts
for 0.77~$\mu\Omega$cm/T$^{1/2}$ and is by a factor of 2 larger
than the \textit{A}=0.38~$\mu\Omega$cm/K$^{1/2}$ coefficient
obtained in constant \textit{B}. Though not yet understood, this
relationship between the negative MR and the occurrence of the
extra --\textit{AT}$^{1/2}$ term in $\rho$(\textit{T}) found for
\textit{T}$\lesssim$12~K and \textit{B}$\geq$1~T (cf. Fig.2) is
very intriguing. Both observations, however, confirm the existence
of a characteristic energy scale of a few K in ThAsSe. We note
that a negative MR is also expected for a 2CK effect originating
from electrical quadrupole moments \cite{Anders}.

Recent examinations of ThAsSe revealed a strong sample dependence
of its resistivity in the high temperature region \cite{cich1}.
This is due to a different amplitude of the increase of
$\rho$(\textit{T}) with decreasing temperature down to around
65~K. Indeed, all the $\rho$(\textit{T}) dependencies can be
mapped to the same
$\frac{\rho(\textit{T})-\rho_{300\textrm{K}}}{\rho_{4\textrm{K}}-\rho_{300\textrm{K}}}$
curve, as demonstrated in Fig. 3(c). Here we compare the
normalized resistivity for the ThAsSe single crystal studied
($\rho_{4 \textrm{K}}/\rho_{300 \textrm{K}}$=1.99) with that one
for the sample characterized by the lowest resistivity ratio
(1.38) measured previously \cite{cich1}. Since our treatment does
not affect the temperature scale, the observed sample dependence
in ThAsSe has clearly a quantitative character only.

The negative temperature coefficient of the resistivity above 65~K
originates, most likely, from a gradual formation of covalently
bonded dimers (As-As)$^{4-}$ \cite{Schoe}. The latter ones have
been recently observed by means of an electron diffraction study
\cite{Withers}. In such a case, the degree of dimerization, being
sensitive to crystal growth conditions, may be responsible for the
varying $\rho_{4 K}/\rho_{300 K}$ values. Nevertheless, the
formation of (As-As)$^{4-}$ does not affect the low-energy physics
in ThAsSe for the following reasons: First, virtually identical
electron diffraction patterns have been obtained at 30 and 100~K
\cite{Withers}. Second, a qualitatively different response of the
resistivity to the application of high pressure has been observed
at low and high temperatures: At 1.88~GPa the maximum of the
resistivity at 65~K is suppressed, whereas the low-\textit{T} term
is completely unchanged \cite{cich1}.

As far as the formation of tunneling centers is concerned, it is,
however, conceivable that some low-energy excitations of singular
(As-As)$^{4-}$ dimers play an important role. A more exciting
possibility concerns the fact that some pnictogen atoms As and
chalcogen atoms Se become involved in a homopolar-to-heteropolar
bond transformation, as discussed for As$_{2}$Se$_{3}$ \cite{Li}
and As$_{2}$S$_{3}$ \cite{Uchino}. The latter scenario is even
more plausible for non-stoichiometric samples \cite{Kniep}. In
other words, we speculate that the movable particle is an
electron, tunneling between As and Se, rather than an atom:
\textit{T}$_{K}$ of a few K in ThAsSe would then be the
consequence of the electron mass being smaller than the atomic
masses by about four orders of magnitude.

In summary, we have investigated a ThAsSe single crystal whose
low-temperature resistivity shows a -\textit{AT}$^{1/2}$ behavior.
Its origin was found to be very different from the frequently
observed quantum interference \cite{Lee, Alts, Bergmann}, as
highlighted by the independence of the resistivity on strong
magnetic fields. Furthermore, the low-\textit{T} thermal
properties give clear evidence for the presence of tunneling
centers in the sample studied. Our experimental findings lead to
the suggestion of a two-channel Kondo effect originating from
interactions between the conduction electrons and TLS. We hope
that our results will have a significant impact on this
interesting field of research, given the fact that the existence,
in real matter, of a two-channel Kondo regime due to tunneling
particles is still a matter of strong current controversy
\cite{Aleiner2, Zarand}.

We would like to express our gratitude to A. Zawadowski for many
enlightening discussions. We are also grateful to A. Bentien, S.
Wirth, and G. Zar\'{a}nd for valuable conversations, and the
Polish Committee for Scientific Research, Grant No. 1 P03B 073 27
(T.C., A.W., and Z.H.) for support. T.C. acknowledges the
Alexander von Humboldt Foundation for a Research Fellowship.


\begin{references}
\bibitem{Kniep}
The chemical composition of various ThAsSe single crystals
(1.5--2~mg each) were determined by the use of the ICP-OES
(Inductively-Coupled-Plasma--Optical-Emission-Spectrometer)
method. The analyses show a slight excess of As in the samples.
\end{references}
\end{document}